\documentclass[aps,prl,twocolumn,groupedaddress]{revtex4}
\usepackage{graphicx}

\begin{document}

%\preprint{}

\title{Thermoelectric response of Fe$_{1+y}$Te$_{0.6}$Se$_{0.4}$: evidence for strong correlation and low carrier density }


\author{A. Pourret$^{1}$, L. Malone$^{2,3}$, A. B. Antunes$^{3}$, C. S. Yadav$^{4}$, P. L. Paulose$^{4}$, B. Fauqu\'{e}$^{1}$ and K. Behnia$^{1}$}
%\thanks{}
\affiliation{(1)Laboratoire de Physique et d'Etude des Mat\'eriaux (ESPCI-UPMC-CNRS),
10 Rue de Vauquelin 75231 Paris, France \\
(2) Laboratoire National des Champs Magn\'etiques Intenses (CNRS-INSA-UJF-UPS), 31400 Toulouse , France\\
(3) Laboratoire National des Champs Magn\'etiques Intenses (CNRS-INSA-UJF-UPS), 38042 Grenoble , France\\
(4) Department of condensed Matter Physics and Material Sciences, Tata Institute of Fundamental Research, Colaba, Mumbai-400005, India}


\date{October 7, 2010}

\begin{abstract}
We present a study of the Seebeck and Nernst coefficients of Fe$_{1+y}$Te$_{1-x}$Se$_{x}$ extended up to 28 T. The large magnitude of the Seebeck coefficient in the optimally doped sample tracks a remarkably low normalized Fermi temperature, which like other correlated superconductors, is only one order of magnitude larger than T$_c$. We combine our data with other experimentally measured coefficients of the system to extract a set of self-consistent parameters, which identify  Fe$_{1+y}$Te$_{0.6}$Se$_{0.4}$ as a low-density correlated superconductor barely in the clean limit. The system is subject to strong superconducting fluctuations with a sizeable vortex Nernst signal in a wide temperature window.

\end{abstract}

%\keywords{}


\maketitle

The discovery of superconductivity in fluorine-doped LaFeAsO\cite{Kamihara} initiated an extensive research activity on iron-based superconductors (for recent reviews see \cite{Ishida,Paglione}). The emergence of a superconducting ground state in proximity to a magnetic instability is a feature common to these systems, cuprate and heavy-fermion superconducors, all suspected to be unconventional. One central subject of investigation is the importance of the electron correlations in the normal state and their possible role in the formation of Cooper pairs. Among these new superconductors, iron chalcogenides with the formula Fe$_{1+y}$(Te$_{1-x}$Se$_{x}$)\cite{Hsu,Sales,Chen} present the simplest crystal structure. Fe$_{1+y}$Te is not a superconductor and undergoes a structural distortion along with the establishment of a long-range SDW order near 65 K. Replacing Te by Se generates a superconducting instability and $T_{c}$ attains a maximum of about 15 K in Fe$_{1+y}$Te$_{1-x}$Se$_{x}$ at x $\simeq$ 0.4. Several recent studies\cite{Tamai,Homes,Aichhorn} have suggested that electronic correlations may be significantly stronger in this family compared to other iron-based superconductors.
%Figure ONE
\begin{figure}
\resizebox{!}{0.6\textwidth}{\includegraphics{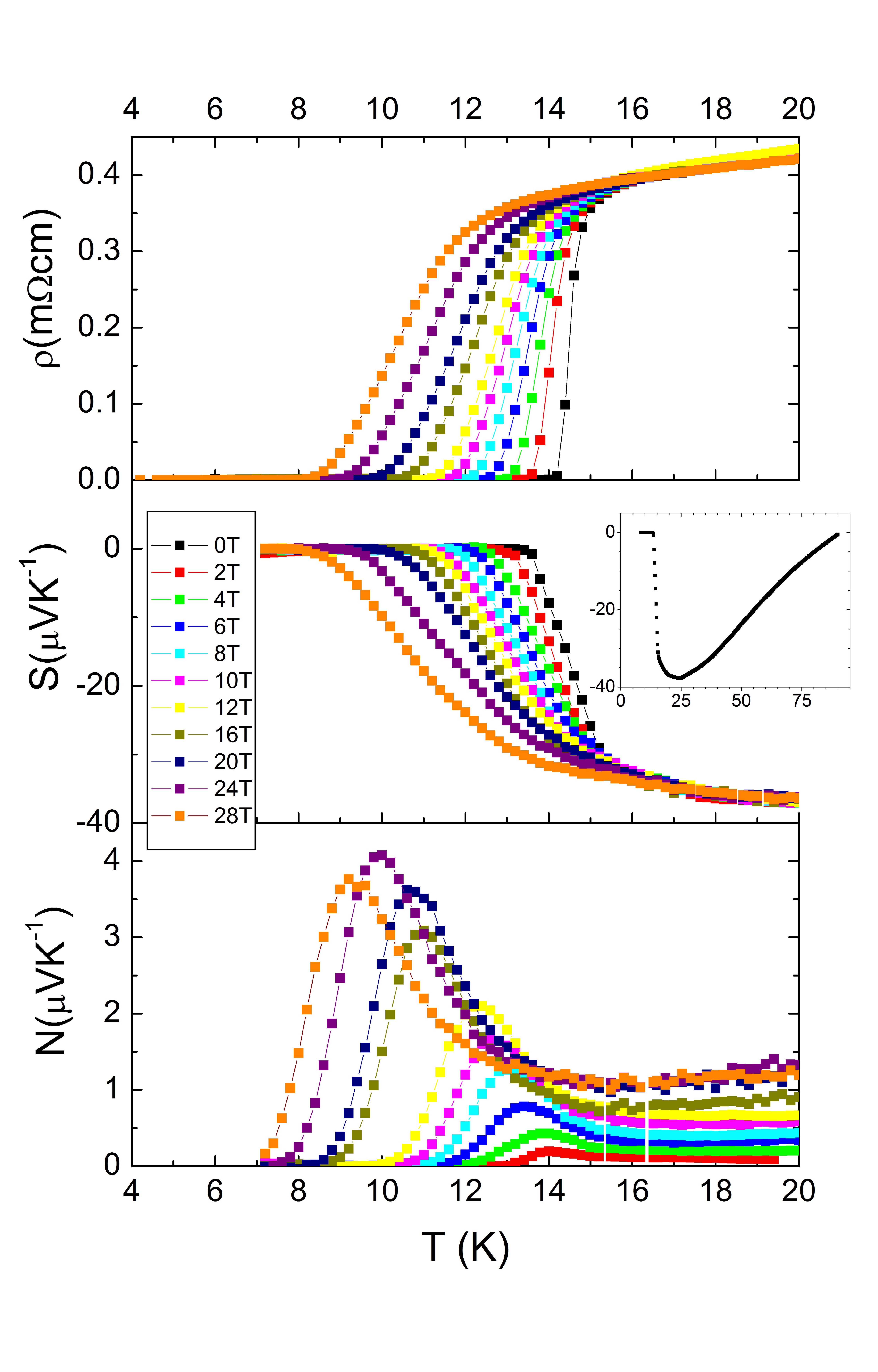}} \caption{\label{fig1}Temperature
dependence of the resistivity (upper panel), the Seebeck coefficient (middle panel),  and
the Nernst coefficient (lower panel) for different magnetic fields in
Fe$_{1+y}$Te$_{0.6}$Se$_{0.4}$. The inset shows the temperature dependence of the zero-field Seebeck coefficient up to 90 K.}
\end{figure}

In this paper, we report on a study of Seebeck and Nernst coefficients of Fe$_{1+y}$Te$_{1-x}$Se$_{x}$ extended to 28 T. We find a large and negative  Seebeck coefficient  in Fe$_{1+y}$Te$_{0.6}$Se$_{0.4}$ and argue that this is a direct consequence of strong electronic correlations leading to a reduced Fermi energy. The Fermi temperature, $T_{F}$, deduced from these measurements yields a large  $T_{c}$/T$_{F}$ ratio comparable to other well-known correlated superconductors. This is a new argument in favor of a magnetically-mediated superconductivity in this system. The determination of $T_{F}$ leads to the extraction of a set of consistent values for carrier density, effective mass, mean-free-path compatible with all known bulk properties of the system. Our analysis identifies Fe$_{1+y}$Te$_{0.6}$Se$_{0.4}$ as a dilute liquid of heavy carriers with a mean-free-path barely exceeding the superconducting coherence length. The Ginzburg number is as large in cuprates, explaining the wide window of thermally-induced vortex flow seen by our Nernst measurements.

Single crystals of Fe$_{1+y}$Te$_{1-x}$Se$_{x}$ were prepared by the chemical reaction of the elements in the stochiometric proportion inside a sealed quartz tube under vacuum described elsewhere\cite{Yadav}. Nernst and Seebeck effects were measured using a one-heater-two-thermometer set-up which allowed us to determine all transport coefficients of the sample in the same conditions. A set-up with cernox thermometers in a He$^{4}$ cryostat was used in a superconducting magnet up to 12 T, and afterwards in a DC resistive magnet  up to 28 T. The magnetic field  was applied
perpendicular to the [applied] heat-current and the [measured] electric-field vectors.

Fig. 1 presents the evolution of the  temperature dependence of transport coefficients near the superconducting transition with magnetic field between 0 T and 28 T in Fe$_{1+y}$Te$_{0.6}$Se$_{0.4}$. The upper panel of Fig 1 presents the in-plane electrical resistivity, $\rho$. The system undergoes a relatively sharp  transition at $T_{c} \sim14$ K in zero magnetic field and the onset of transition is barely affected by the magnetic field. The middle panel shows the temperature dependance of the Seebeck coefficient, $S$, at different magnetic fields. As in other iron-based superconductors\cite{CWang}, $S$ is large. It peaks to  ($S\sim$ -38 $\mu$VK$^{-1}$) around $T\sim$ 22 K (See the inset), the largest absolute value reported in the Fe$_{1+y}$Te$_{1-x}$Se$_{x}$ family\cite{Pallecchi}. The application of a magnetic field shifts and broadens the superconducting transition, but not the amplitude of the normal state Seebeck response.

The lower panel shows the field and temperature dependence of the Nernst signal, $N = -E_{y}/\Delta_{x}T$. As seen in the figure, $N$ is markedly enhanced in the vortex liquid state and vanishes with the solidification of the vortex lattice. Its peak shifts to lower temperature with increasing magnetic field. These features are reminiscent of previous reports on thermally-induced motion of vortices in cuprate\cite{Ri,YWang}, organic\cite{Nam} and conventional\cite{Pourret} superconductors. In the vortex liquid state, vortices move under the influence of a thermal gradient because of the excess of entropy of their cores, generating a transverse voltage. The amplitude of the Nernst peak steadily increases up to 24 T and begins to decrease afterwards when the overlap between the vortex cores becomes sufficiently large to compensate for the increase in the number of vortices.

Fig. 2 presents contour plots of the Seebeck and the Nernst $\nu = -E_{y}/ B \Delta_{x}T$ coefficients in the (B,T) plane. Several previous works\cite{Kida,Braithewaite,Fang} have reported on the strikingly large slope of the upper critical field at $T_c$ ($\frac{dHc2}{dT}|_{T_{c}}$). For a magnetic field along the c-axis, the field-dependence of the midpoint of the resistive transition yields a value significantly lower than what can be extracted from the jump in specific heat\cite{Braithewaite,Klein}.  As seen in the upper panel of the figure, the onset of the drop in $S/T$ yields a slope of 10 T/ K, slightly lower than what was very recently deduced from specific heat measurements by Klein \emph{et al.}\cite{Klein} (12 T/K). As seen in the lower panel of Fig.2, in a wide window of temperature and magnetic field, the Nernst response is enhanced above its normal-state value revealing an extended fluctuating regime with no true thermodynamic phase transition separating the vortex liquid and the normal state.

%Figure Two
\begin{figure}
\resizebox{!}{0.6\textwidth}{\includegraphics{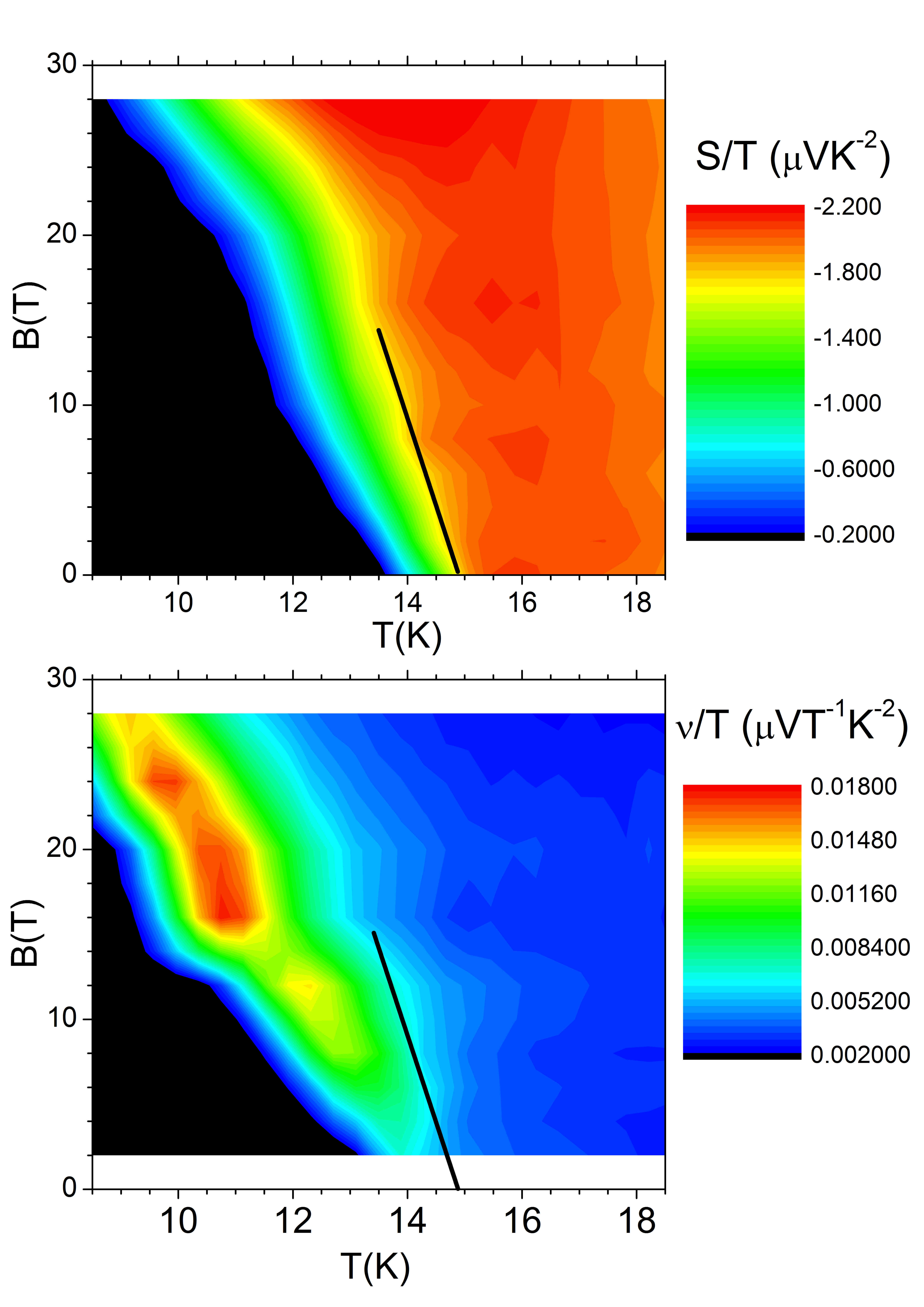}} \caption{\label{fig2} Contour plot of $S/T$ (top) and $\nu/T$ (bottom), in the (B,T) plane.  In the vortex solid state (the black region), the electric field vanishes. Because of the extreme sensitivity of the Nernst effect to the vortex motion, its size is smaller in the lower plot. Note the wide fluctuation region close to $T_{c}$.  The thick line in both panels point to a slope of 10 K/T slightly lower than $\frac{dHc2}{dT}|_{T_{c}}=12 K/T$ according to specific heat measurements\cite{Klein}.}
\end{figure}

We now turn our attention to the amplitude of the Seebeck coefficient. Fig. 3 presents the temperature dependence of the Seebeck coefficient divided by temperature in Fe$_{1+y}$Te$_{1-x}$Se$_{x}$ with different Se content. The zero-temperature extrapolated value of $S/T$ increases with Se content, starting from $S/T$=-0.4 $\mu$V/K$^{2}$ for non-superconducting FeTe and reaching $S/T$=-2.9 $\mu$V/K$^{2}$ for the optimally doped compound Fe$_{1+y}$Te$_{0.6}$Se$_{0.4}$. We checked our incertitude on the magnitude of $S/T$ by measuring several optimally-doped samples. As seen in the inset of Fig.3, our data on three different samples set $S/T$ in $T\longrightarrow 0$ limit to be 2.8 $\pm 0.3 \mu$V/K$^{2}$.

Diffusive Seebeck response of a Fermi liquid is expected to be T-linear in the zero-temperature limit, with a magnitude proportional to the strength of electronic correlations as in the case of the T-linear electronic specific heat, $C_{e}/T=\gamma$. Both of them can be linked to the Fermi temperature, $T_{F}$:

\begin{equation}\label{1}
S/T= \pm\frac{\pi^{2}}{2}\frac{k_{B}}{e}\frac{1}{T_{F}}
\end{equation}
\begin{equation}\label{2}
 \gamma= \frac{\pi^{2}}{3}k_{B}\frac{n}{T_{F}}
\end{equation}
where $k_{B}$ is Boltzmann's constant, $e$ is the electron charge, and $n$ is the carrier density. In a multi-band system with both electrons and holes contributing with opposite signs to the overall Seebeck response, this one-band formula sets an \emph{upper limit} to the Fermi temperature of the dominant band. In a wide range of half-filled correlated metals with a carrier density of one carrier per formula unit, the magnitude of  $S/T$ correlates with $\gamma$\cite{Behnia1}.

Here one can deduce $T_{F}$=151 K from $S/T$= 2.8 $\mu V/K^{2}$, implying a ratio of the critical temperature to the normalized Fermi temperature as large as 0.1. For two other superconductors believed to be unconventional, namely  CeCoIn$_{5}$ (12 $\mu V/K^{2}$)\cite{Izawa} and  YBa$_{2}$Cu$_{3}$O$_{6.67}$ (-0.4 $\mu V/K^{2}$)\cite{Chang}, one can insert in Eq.1 the reported magnitude of $S/T$ and extract $T_{F}$ in a similar way. As seen in Fig. 4,  $\frac{T_{c}}{T_{F}}$ ratio is of the same order of magnitude in the three systems. This figure is a plot first introduced by Moriya and Ueda\cite{Moriya} suggesting an intimate link between $T_c$ and the coherence temperature of a correlated electron system when superconductivity is mediated by spin fluctuations. As in the case of PuGaIn$_5$\cite{Sarrao}, this observation argues in favor of superconductivity mediated by electron correlations in Fe$_{1+y}$Te$_{0.6}$Se$_{0.4}$.

% Figure Three
\begin{figure}
\resizebox{!}{0.38\textwidth}{\includegraphics{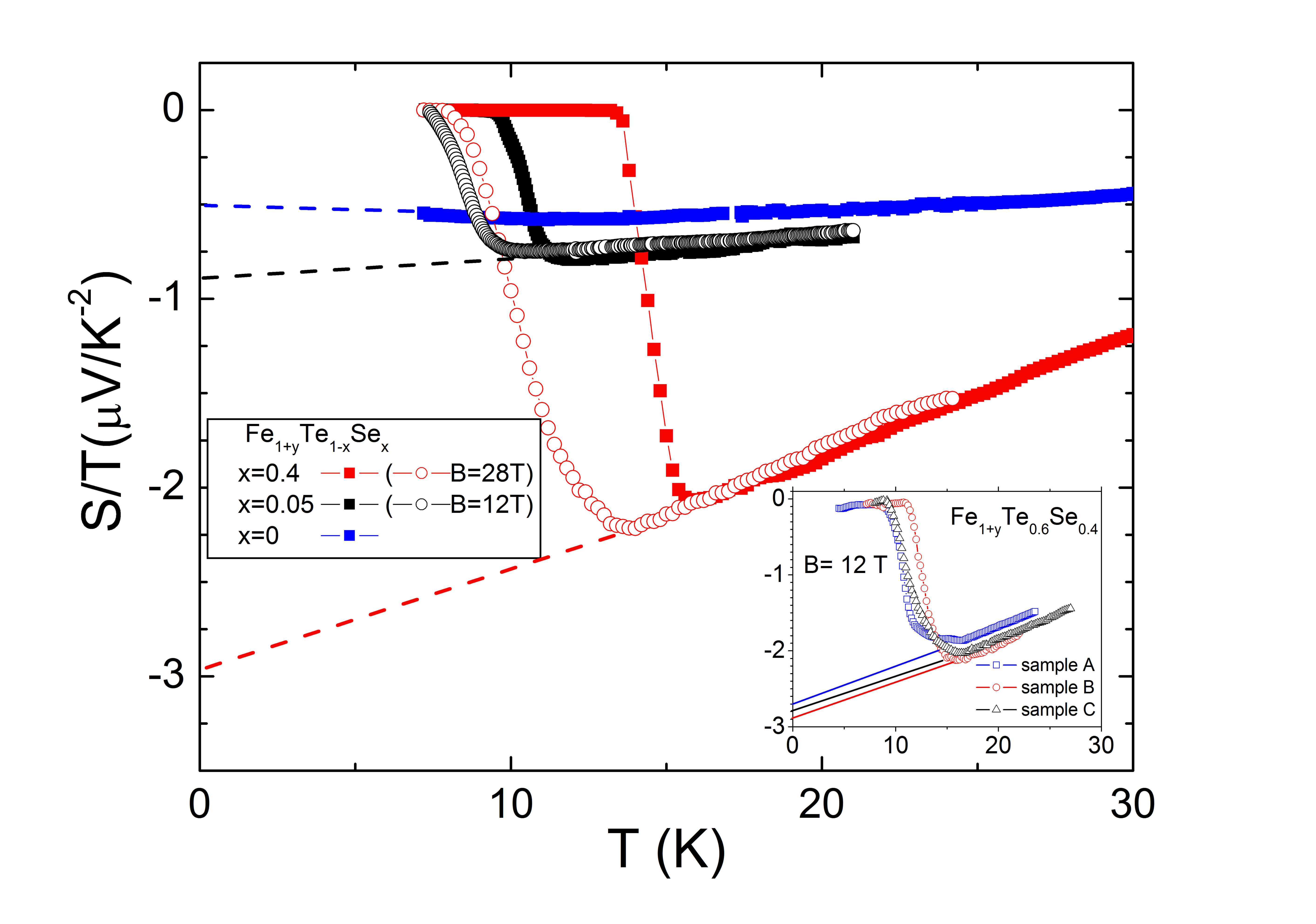}} \caption{\label{fig3}Temperature
dependence of the Seebeck coefficient divided by T, $S/T$ in Fe$_{1+y}$Te$_{1-x}$Se$_{x}$, with x=0,0.05 and 0.4. In the superconducting samples a magnetic field was applied to partially recover the normal state. Inset presents the temperature dependence of $S/T$ for three optimally-doped samples at B=12 T.}
\end{figure}

%Figure Four
\begin{figure}
\resizebox{!}{0.35\textwidth}{\includegraphics{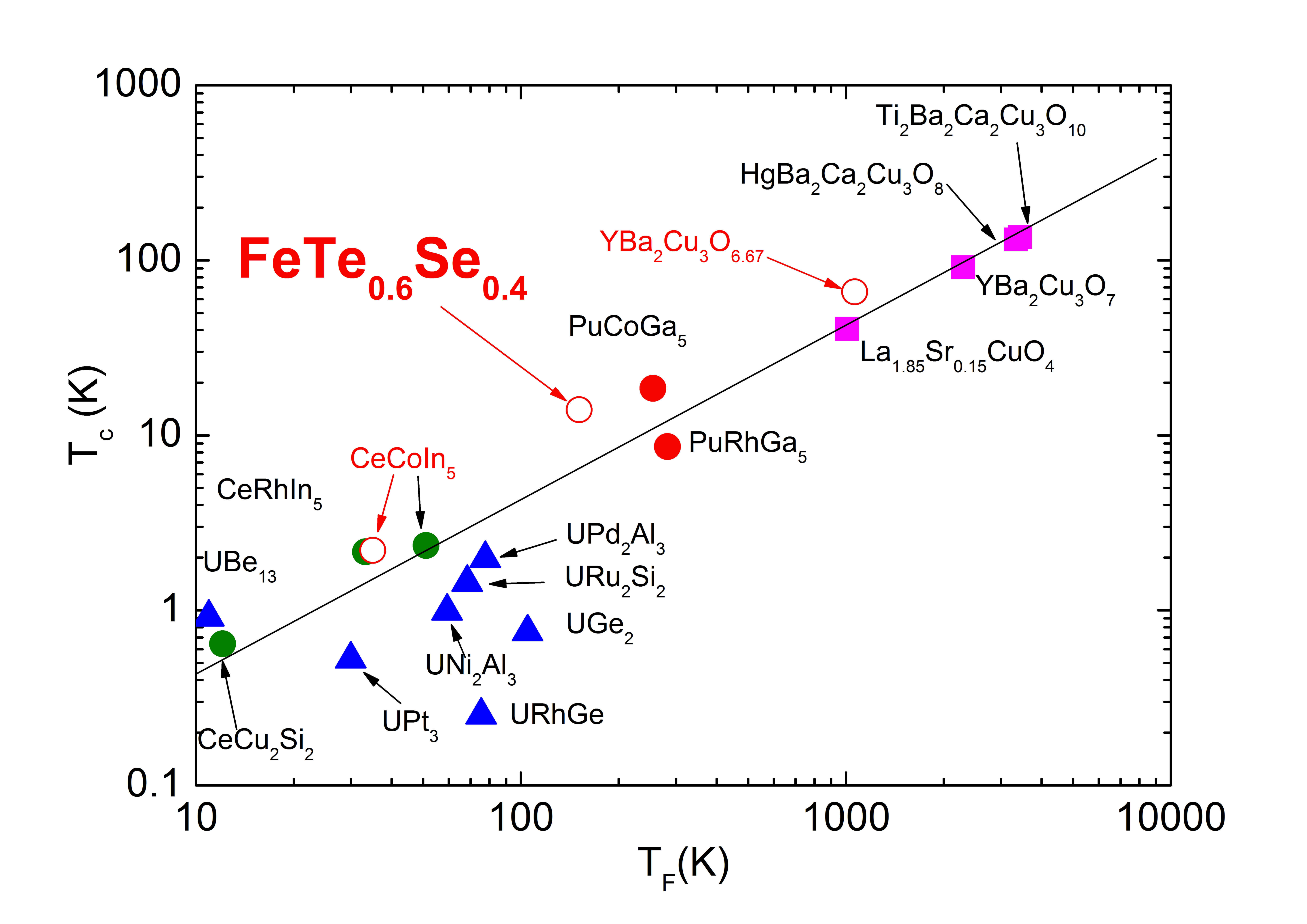}} \caption{\label{fig4} The Morya-Ueda plot : $T_{c}$, as a function of  Fermi temperature in a number of unconventional superconductors\cite{Sarrao}. The three red open circles represent three superconductors, for which T$_{F}$ was extracted from $S/T$ using Eq. 1 [See text].}
\end{figure}

 Note that the low value of $T_{F}$ or the strength of electron correlations cannot be deduced from the magnitude of $\gamma$ alone. In Fe$_{1+y}$Te,  $\gamma$ can be measured down to low temperatures and is reported to be 34 mJ/molK$^{2}$\cite{Chen}. In the optimally doped Fe$_{1+y}$Te$_{0.6}$Se$_{0.4}$, two recent studies\cite{Klein, Tsurkan} find  $\gamma$= 23$\pm$ 3 mJ/molK$^{2}$, a value significantly lower than what was initially reported\cite{Sales}. The discrepancy is mostly due to the difficulty to extract the lattice contribution, which by far dominates the total specific heat at T$_{c}$.  Now as seen above and detailed in table I, the magnitude of $S/T$  is an order of magnitude \emph{larger} in the optimally doped system. The absolute value of the dimensionless ratio of thermopower to specific heat ($q=\frac{N_{Av}e S}{T\gamma}$; where N$_{Av}$ is the Avaogadro number)\cite{Behnia1},  is close to unity in the undoped system, but approaches 12 in the optimally-doped case. This means that while Fe$_{1+y}$Te is half-filled, in Fe$_{1+y}$Te$_{0.6}$Se$_{0.4}$ the Fermi surface occupies only about 0.04 of the volume of the Brillouin zone. Thus, while the Density Of States (DOS) \emph{per volume} is lower in the optimally-doped compound, the DOS \emph{per carrier} is much larger.

 To underline what is striking about Fe$_{1+y}$Te$_{0.6}$Se$_{0.4}$, table I compares it with the borocarbide  LuNi$_{2}$B$_{2}$C\cite{Rathnayaka,Michor}, a conventional superconductor with a similar $T_c$. Both $\gamma$ and the superconducting gap are comparable in the two systems\cite{Hanaguri,Dewilde}. But  $\frac{dH_{c2}}{dT}|_{T_{c}}$ and $S/T$ are 15 to 20 times larger in the iron-based superconductor, which has a lower Fermi energy and a shorter coherence length, both drastically reduced by the combination of mass enhancement and carrier density reduction.

%Table1
\begin{table}
\begin{tabular}{|l||c|c|c|r|}
\hline
\hline
Parameter& Fe$_{1+y}$Te & Fe$_{1+y}$Te$_{0.6}$Se$_{0.4}$ & LuNi$_{2}$B$_{2}$C \\
\hline \hline
$\gamma$ (mJ/molK$^{2}$)& 34  & 23$\pm$3 & 19 \\
\hline
S/T  ($\mu$V/K$^{2}$)&-0.4  & -2.8$\pm$0.3 & -0.22\\
\hline
q  (=$\frac{SN_{Av}e}{T\gamma}$)&-1.1  & -11.7 & -1.1\\
\hline
T$_{c}$ (K)& -- & 14 & 16 \\
\hline
$\Delta_0$ (meV)& --  & 1.7  & 2.2 \\
\hline
$\frac{dH_{c2}}{dT}|T_{c}$ (T/K)& --  & 12$\pm$2 & 0.6 \\
\hline \hline
\end{tabular}
\caption{\label{table1}Various physical properties of Fe$_{1+y}$Te$_{0.6}$Se$_{0.4}$ compared with the undoped system and a borocarbide superconductor with a similar T$_{c}$. }
\end{table}

Let us conclude by checking the quantitative consistency of this analysis. According to the  thermopower-to specific heat ratio, the carrier density is $|q|^{-1}$ = 0.085 carrier per unit cell. Given the volume of the latter ($\sim 0.078$ nm$^{-3}$\cite{Hsu}), this yields $n= 1.1\times$10$^{21}$ cm$^{-3}$, which sets $k_F= (3\pi^{2} n)^{1/3} $. Combining the deduced value of $k_F$ with T$_{F}$=151 K leads to the effective mass, m$^{*}$ and the Fermi velocity, v$_F$ by using two simple equations k$_{B}$T$_{F}=\frac{\hbar^{2}k_F^{2}}{2m^{*}}$ and $\hbar k_F=m^{*}v_{F}$. The mean-free-path, $\ell$ can also be estimated from  $k_F$ and the measured resistivity ($\rho_{0} \sim$ 0.35 m$\Omega$ cm) using $\rho_{0}^{-1}=\frac{2}{3\pi}\frac{e^{2}}{h}k_F^{2}\ell$. The superconducting coherence length is set by the the slope of the upper critical field at T$_c$: $\xi^{-2}=\frac{2\pi}{\Phi_{0}}0.69\frac{dH_{c2}}{dT}|_{T_{c}}$. The results are listed in table II.

To check the consistency of these parameters, let us note that in a BCS superconductor, $\xi$, v$_F$ and $\Delta_0$ are related by the equation $\xi=\frac{\hbar v_F}{\pi \Delta_{0}}$. The measured $\Delta_{0}$ (1.7meV) and the estimated v$_F$ (1.2 10$^4$ m/s), would yield $\xi\simeq 1.6$ nm, close to what is directly extracted from the slope of the upper critical field\cite{Klein}. These values are in rather good agreement with the conclusions of a recent ARPES study\cite{Tamai}.

%Table2
\begin{table}
\begin{tabular}{|l||c|c|r|}
  \hline
Quantity & Magnitude \\
\hline \hline
$k_F$ (nm$^{-1}$)& 3.2  \\
\hline
$\ell$(nm)& 3.4 \\
\hline
$\xi$(nm)& 1.6 \\
\hline
m$^{*}$ (m$_{e}$) & 29\\
\hline
v$_{F}$ (km/s)& 12 \\
\hline \hline
\end{tabular}
\caption{\label{table2} Set of parameters for Fe$_{1+y}$Te$_{0.6}$Se$_{0.4}$ compatible with all measured bulk properties. See the text for details of their derivations.}
\end{table}

The success of the simple one-band picture used here  suggests that the superconducting and normal properties are both dominated by a single electron-like band. Thus, Fe$_{1+y}$Te$_{0.6}$Se$_{0.4}$ is a correlated metal in its normal state with a low density of heavy quasi-particles and barely clean in its superconducting state. As seen in Table II, the inequality $\ell > \xi$ is \emph{in extremis} respected. Remarkably, the average distance between electrons, the average size of a Cooper pair and the electron mean-free-path are, all three, of the same order of magnitude but respect the hierarchy required for unconventional superconductivity, which would be destroyed if $\ell$ falls shorter than $\xi$. The fragility of this hierarchy and its possible breakdown may be a clue to the absence of bulk superconductivity in a wide $x$ window in the Fe$_{1+y}$Te$_{1-x}$Se$_{x}$\cite{Sales}. On the other hand, the very short coherence length and the small condensation energy per volume yield a Ginzburg number as large as in cuprates(10$^{-2}$)\cite{Blatter}, providing a natural explanation for the wide flux flow window detected by our Nernst measurements.

In summary, we measured the field and temperature dependence of the Nernst and Seebeck coefficients in Fe$_{1+y}$Te$_{0.6}$Se$_{0.4}$ and determined the Fermi temperature. The $T_c/T_F$ ratio is as large as in any known superconductor. We are grateful to Jacques Flouquet for introducing us to the Moriya-Ueda plot and to Thierry Klein and Zengwei Zhu for stimulating discussions. This work is supported by the Agence Nationale de la Recherche as a part of DELICE project (ANR-08-BLAN-0121-02) and by EuroMagNET II under the EU contract number 228043.

\end{document}